\documentstyle[12pt]{article}

\begin{document}



\begin{center}
\bigskip
{\bf Localized Structures of Electromagnetic Waves in Hot 
Electron-Positron Plasma }
\vskip2truecm
{ S. Kartal,\footnote{\normalsize Permanent address:
University of Istanbul, Department of Physics, 34459, Vezneciler-Istanbul,
Turkey.} L.N. Tsintsadze,\footnote{\normalsize Present address: Faculty of
Science, Hiroshima University, Hiroshima 739, Japan. 

Permanent address: Institute of
Physics, The Georgian Academy of Science, Tbilisi 380077,  Republic
of Georgia} and V.I. Berezhiani}\\
{International Centre for Theoretical Physics, Trieste, Italy.}
\end{center}
\vskip1truecm
\centerline{ABSTRACT}
\bigskip

The dynamics of relativistically strong electromagnetic (EM) wave propagation
in hot electron-positron plasma is investigated. The possibility of finding
localized stationary structures of EM waves is explored. It is shown that
under certain conditions the EM wave forms a stable localized soliton-like
structures where plasma is completely expelled from the region of EM field
location. 
\\ \\
PACS number(s): 52.60.+h, 52.40.Db

\newpage
\null
\baselineskip=22pt 
During the last few years a considerable amount of work has been 
devoted to the analysis of nonlinear electromagnetic (EM) wave 
propagation in electron-positron (e-p) plasmas [1]. Electron- positron pairs 
are thought to be a major constituent of the plasma emanating both from 
the pulsars and from inner region of the accretion disks surrounding the 
central black holes in the active galactic nuclei (AGN) [2]. The process of 
e-p pair creation occurs in relativistic plasma at high temperatures, 
when the temperature of the plasma becomes of the order of, or larger than,
the rest energy of electrons. Such relativistic plasmas have presumably
appeared in the early universe [3]. Intense relativistic e-p 
plasmas could also exist in the vicinity of cosmic defects like 
superconducting cosmic strings [4]. Collective processes in e-p plasmas 
are of considerable interest. The processes of wave self-modulation of EM 
waves and soliton formation have attracted a great deal of attention. 
Stable localized solution may be a potential mechanism for the production 
of micro-pulses in AGN and pulsars [5]. In the early universe  stable 
localized EM waves could create inhomogeneities necessary to understand 
the observed structure of the visible universe.

In the recent paper of Berezhiani and Mahajan [6] the nonlinear propagation of relativistically 
strong EM radiation in a hot e-p plasma has been considered. It has been 
shown  that e-p plasma supports the propagation of nondiffracting and 
nondispersive EM pulses (light bullets) with large density bunching. 
However, the authors concentrated in the case of transparent plasma and 
consequently the group velocity of the pulses is close to the velocity of light 
c.
In the present paper we consider the propagation of  
strong EM radiation in a hot unmagnetized  e-p plasma aiming to find the 
localized stationary soliton-type solutions. 

We start from Maxwell equations to describe the EM wave propagation in an 
e-p plasma, expressing the fields by the vector and scalar potentials, i.e.

$$ {\bf E}=-{1 \over c}{\partial {\bf A}\over \partial t} - \nabla 
\varphi ~~, ~~~~~~~~~~~{\bf B}=\nabla \times {\bf A}   \eqno (1) $$
where the coulomb gauge $ ~{\nabla \cdot {\bf A}=0}~ $ is fulfilled. 
Accordingly the field equations take the form:

$${ \partial ^{2}{\bf A}\over \partial t^{2}}- c^{2}\Delta {\bf A}+ c{ \partial \over  
 \partial t}(\nabla \varphi )- 4 \pi c {\bf J} =0   \eqno (2) $$
and

$$ \Delta \varphi= -4 \pi \rho  ~~~~~~~~~~~~~~~~~~~~~~~~~ \eqno (3) $$
Here, $\rho $ and ${\bf J}$ are the charge and current densities given by

$$  \rho= \sum_{\alpha} e_{\alpha}n_{\alpha},~~~~~~~~~~~ {\bf J}=\sum_{\alpha} 
e_{\alpha}n_{\alpha}{\bf u}_{\alpha} ~~~\eqno  (4)$$
where $\alpha $ indicates the particle species $\alpha$ ($ = e, p $ for 
electrons and 
positrons, respectively); $ e_{\alpha }$ and $ n_{\alpha }$ are the charge and 
density of the corresponding particle $\alpha $. We consider the case in which the 
equilibrium state is characterized by $ n_{0e}=n_{0p}=n_{0}$, where $n_{0 \alpha}$ 
is the equilibrium density of the particle $\alpha $.

Before writing the hydrodynamic equations of relativistic plasma it is necessary to 
define what relativistic means. In fact, we have two types of relativistic regimes 
in plasma: 
in a strong EM field the plasma particles may obtain relativistic velocities. In 
space, the EM radiation of objects (nuclei of galaxies, radio-galaxies, quasars, 
pulsars, etc.) may serve as a source of such strong fields. The case when the 
thermal energy of the plasma particles is of the order of, or larger than, the energy 
at rest, it is the other type of relativistic regime. In this case the thermal velocities 
of the particles become of the order of the light speed. Certainly, in both cases 
the decisive role belongs to relativistic effects in plasma, but the character of 
its manifestation is different. Both relativistic effects can play an important 
role in the e-p plasma. Let us assume that the velocity distribution of the 
particles 
of species $\alpha$ is locally a relativistic Maxwellian. Then, according to 
Ref.[6, 7], the set of relativistic hydrodynamic equations of motion can be written as:

$$ {d \over dt }(m_{0\alpha }G_{\alpha}\gamma _{\alpha} c^{2})- {1 \over 
n_{\alpha}}{\partial \over \partial t}{P_{\alpha}}=e_{\alpha} 
{\bf u}_{\alpha}{\bf E} ~~~~~~~  \eqno (5) $$

$$ {d \over dt}({\bf p}_{\alpha}G_{\alpha})+{1 \over n_{\alpha}}\nabla P_{\alpha}= 
e_{\alpha}{\bf E}+{e_{\alpha} \over c}({\bf u_{\alpha}\times B})~~~~~ \eqno (6)  $$

The continuity equation for the particle $\alpha $ is

$${\partial n_{\alpha}\over \partial t}+\nabla (n_{\alpha}{\bf 
u_{\alpha}})=0   \eqno (7)  $$
Here $~{\bf p_{\alpha}}=\gamma_{\alpha}m_{0 \alpha}{\bf u}_{\alpha}~$ is the 
hydrodynamic momentum, $~P_{\alpha}=n_{\alpha}T_{\alpha}/\gamma_{\alpha}~$ is the 
relativistic particle pressure, ${\bf u_{\alpha}}$ is the hydrodynamic 
velocity of the 
fluid, $~ \gamma_{\alpha} = (1- u_{\alpha }^{2}/c^{2})^{-1/2} ~ $ is the 
relativistic factor, $m_{0 \alpha}$ and $ T_{\alpha}$ are the particles 
invariant rest mass, and temperature respectively, $~ 
d_{\alpha}/dt=\partial /\partial t + {\bf u}_{\alpha}\nabla ~$ is the comoving 
derivative. The role of the particle-mass is now played by the quantity $~ 
M_{eff}=m_{0\alpha}G_{\alpha}(z_{\alpha})~$, where 
$~G_{\alpha}(z_{\alpha})=K_{3}(z_{\alpha})/K_{2}(z_{\alpha})~$. Here 
$K_{3}(z_{\alpha})$ and $K_{2}(z_{\alpha})$ are respectively the McDonald 
functions of the second and third order $~ 
(z_{\alpha}=m_{0\alpha}c^{2}/T_{\alpha})~$. The effective mass of the particles 
$M_{eff}$ depends on the temperature. For nonrelativistic temperatures 
$~(T_{\alpha} \ll m_{0\alpha}c^{2})~$ 
$~M_{eff}=m_{0\alpha}(1+5T_{\alpha}/2m_{0\alpha}c^{2})~$, while for  
ultrarelativistic high temperatures $~(T_{\alpha} \gg m_{0\alpha}c^{2})~$ the 
effective mass becomes $~M_{eff}=4T_{\alpha}/c^{2}~$, and the fluid 
inertia is 
primary provided by random thermal motion of the particles. In this case, the 
rest-mass is negligibly small and the e-p gas behaves like photons. Using simple 
manipulation, from the Eqs. (5) and (6), we obtain the adiabatic equation 
which reads

$$ {n_{\alpha} z_{\alpha}\over \gamma _{\alpha}K_{2}(z_{\alpha})}exp( 
- z_{\alpha}G_{\alpha}(z_{\alpha}))=const.    \eqno(8) $$

In the nonrelativistic limit $(z_{\alpha}\gg 1)$ , Eq. (8) yields the result 
for a mono 
atomic ideal gas $~(n_{\alpha}/(\gamma _{\alpha}T_{\alpha}^{3/2})=const.)~$, and in 
the ultrarelativistic limit $(z_{\alpha}\ll 1)$ we have the adiabatic law for 
the "photon" gas $(n_{\alpha}/(\gamma _{\alpha}T_{\alpha}^{3})=const.)$.

We are looking for a localized one-dimensional solution of this system of 
equations for a circularly polarized EM wave, where the vector potential 
${\bf A}$ can be expressed as:

$$ {\bf A_{\bot}}={1 \over 2}({\bf {\hat x}}+i{\bf {\hat y}})A(z,t)exp 
(-i {\omega_{0}}t)+c.c  ~~~~\eqno (9) $$
where $ A(z,t)$ is a slowly varying function of $t$, $\omega_{0}$ 
is mean frequency , ${\bf {\hat x}}$ 
and ${\bf {\hat y}}$ are the unit vectors. The transverse component of 
equation of motion (6) is immediately integrated to give (for details 
see Ref. [6]):

$$ {\bf p_{\alpha \bot}}G_{\alpha}=-{e_{\alpha}\over c}{\bf A_{\bot}} 
\eqno (10) $$
where the constant of integration is set equal to zero, since the 
particle hydrodynamic momentum is assumed to be zero at infinity where 
the fields vanish.

Before writing the equations for the longitudinal motion we would like 
to mention that this motion of the plasma is driven by the 
ponderomotive
pressure ($\sim p_{\alpha \bot}^{2}$) of high-frequency EM fields and it 
does not depend on the particle charge sign. In what follows we assume 
that in  equilibrium the temperatures of electrons and positrons are 
equal, i.e. $T_{0e}=T_{0p}=T_{0}$. Since the effective mass of the 
electrons and positrons are equal ($G_e=G_p=G$), the radiation pressure 
gives  equal longitudinal momenta to both the electrons and positrons 
($p_{ez}=p_{pz}=p_z$) and affects 
concentration without producing the charge separation. Consequently $ 
n_{e}=n_{p}=n $ and $\phi =0 $. It is also evident that due to symmetry 
between electron and positron fluids their temperatures remain equal
$(T_{e}=T_{p}=T)$. 

It is now convenient to introduce the following dimensionless quantities:

$$ {\bf p_\alpha}={{\bf p_\alpha }\over m_{0e}},~~ n={n \over n_0},~~
T={T \over m_{0e}c^2},~~{\bf A}={|e|{\bf A}\over m_{0e}c^2},~~
{\bf r}={\omega_{e}\over c}{\bf r},~~ t=\omega_e t   \eqno(11)$$
where $\omega_e=(4\pi e^2n_0/m_{0e})^{1/2}$ is the electron Langmuir
frequency.

The longitudinal motion of the plasma is determined entirely by the set 
consisting of the $z$ component of the equation of motion (6),

$$\left( {\partial\over \partial t}+u_z{\partial\over \partial z}\right)
Gp_z+{1\over n}{\partial\over \partial z}{nT\over \gamma}=
- {1\over 2\gamma G}{\partial |A|^2\over \partial z}   \eqno(12)$$
and the "energy" conservation equation (5),

$$\left( {\partial\over \partial t}+u_z{\partial\over \partial z}\right)
G\gamma-{1\over n}{\partial\over \partial t}{nT\over \gamma}=
{1\over 2\gamma G}{\partial |A|^2\over \partial t}  \eqno(13)$$
where $u_z=p_z/\gamma$. The relativistic factor $\gamma$ does not depend 
on the "fast" time ($\omega_0^{-1}$) and can be written as:

$$\gamma=\left[1+{|A|^2\over G^2}+p_z^2 \right]^{1/2}  \eqno(14)$$

Substituting (9) and (10) into Eq.(2), then for a slowly varying 
amplitude of EM wave $A(z,t)$ we obtain the following equation:

$$2i\omega_0{\partial A\over \partial t}+{\partial^2 A\over \partial z^2}+
\Delta\cdot A+2 f A=0   \eqno(15)$$
where 
$$f=1-{nG_0(T_0)\over \gamma G(T)}   \eqno(16)$$
and $\Delta=\omega_0^2-2$. For convenience we redefined the electron rest 
mass in Eq.(11) as $m_{0e}\rightarrow m_{0e}G_0(T_0)$. In dimensional units
$\Delta\sim \omega_0^2-2\omega_e^2$, where 
$\omega_e=(4\pi e^2 n_0/m_{0e}G_0(T_0))^{1/2}$. 

We are looking for the stationary localized solutions (vanishing at 
infinity) 
described by Eqs. (8), (12)-(16). Assuming that $|A|$ depends only on the 
special coordinate $z$, and integrating Eqs. (12) and (13), we get the 
following integral of motion

$$G(T)\gamma=G_0(T_0)  \eqno (17)$$
where $\gamma=(1+|A|^{2}/G^{2})^{1/2} $ ($p_z=0$). From Eq. (17) we get:

$$G=G_0 \left (1-{|A|^2 \over G_0^2}\right )^{1/2}  \eqno (18)$$

It follows from Eq.(18) that the present hydrodynamical theory, which 
describes the nonlinear waves in e-p plasma, is valid for 
$|A|_{max}^{2}/G_0^2 \leq 1$. When the latter is violated, then the 
electromagnetic waves are overturned and they will cause multi-stream 
motion of the plasma. In such a situation, one must resort to kinetic 
description for studying the nonlinear wave motion. This investigation 
is, however, beyond the scope of the present paper. 

Using Eqs. (17)-(18) 
we get $f=1-n$. Now we should obtain a relationship between $n$ and $|A|^2$. 
To this end, one can use the adiabatic equation (8). Unfortunately it is 
impossible to solve this problem analytically for the arbitrary temperatures. 
In the nonrelativistic case $( T,T_0 \ll 1)$  Eq. (8) gives 
$n=\gamma(T/T_0)^{3/2}$ and using Eqs. (17)-(18) along with the asymptotic 
expression for $G$ ($=1+{5 \over 2}T$) for the plasma 
density we obtain:

$$ n=\left ( 1- {|A|^2\over 5T_0}\right )^{3/2}  \eqno (19)$$
In the ultrarelativistic case $(T,T_0\gg 1)$ 
from Eq. (8) we have $n=\gamma (T/T_0)^3$ 
and using the asymptotic expression for $G$ ($=4T$) we get:

$$ n=1-{|A|^2 \over 16T_0^2} \eqno (20) $$
The expressions (19) and (20) show that the total density of plasma can 
become zero at $|A|^2=|A|_{cr}^2$ (where $|A|_{cr}^2=5T_0$ for the 
nonrelativistic case  and $|A|_{cr}^2=16T_0^2$ for the 
ultra-relativistic case). This phenomenon can be called 
"electron-positron cavitation" and for the ultrarelativistic case has been 
discovered in Ref. [8]. It should be noted that the upper bound limitation of 
the amplitude of the vector potential is caused by the fact that we 
consider here the stationary case, and consequently the inertial terms in 
Eqs.(12)-(13) have been neglected. In the case when $|A|^2>|A|_{cr}^2$, 
a gas-dynamic pressure force cannot compensate the ponderomotive one and 
a stationary distribution of fields does not exist. 

In the ultrarelativistic case, $f=|A|^2/16T_0^2$ and Eq. (15) takes the 
form of the well known nonlinear Schr\"{o}dinger equation. The stationary 
soliton solution of this equation (which corresponds to nonlinear 
frequency shift $\Delta =-{A_m^2 / 16T_0^2}$) is 
$$ A=A_m sech \left ( {A_m \over 4T_0}z \right ) \eqno (21)$$
were $A_m$ is amplitude of soliton. Note that amplitude of the soliton 
$A_m$ can be 
relativistically strong $(A_m \gg 1)$. The only restriction is that $A_m \leq 
A_{cr} =4T_0$. If $ A_m \rightarrow A_{cr}$ then the cavitation of plasma 
occurs and all particles are rejected from the central part of soliton. 

Now let us consider the nonrelativistic case. Using Eq. (19) $f$ can be 
written as
$$ f=1-\left(1-{|A|^2\over 5T_0}\right )^{3/2} \eqno(22)$$
substituting Eq.(22) into Eq.(15) we get the nonlinear Schr\"{o}dinger 
equation with a saturating nonlinearity. For the stationary soliton 
solution we should solve to following equation:
$${d^2 E \over dz^2}-\lambda^2 E +2 E 
[1-(1-E^2)^{3/2}]=0 \eqno(23) $$
where $E=|A|/(5T_0)^{1/2}\leq 1$ and $\Delta =-\lambda^2$. The 
Eq.(23) has a soliton solution provided that nonlinear frequency shift 
satisfies the following "dispersion" relation:
$$ \lambda^2=2-{4\over 5}{1\over E_m^2}\left ( 
1-(1-E_m^2)^{5/2}\right) \eqno(24) $$
where $E_m$ is the amplitude of the soliton. One can see that $\lambda^2$ 
monotonically grows with $E_m^2$ and when $E_m^2\rightarrow 1$ 
(which corresponds to cavitation) obtains its maximal allowed value 
$\lambda_{max}^2=1.2$. Unfortunately the general analytical solutions of 
Eq.(23) can not be expressed in terms of elementary functions. We would 
like to mention that in the case of small amplitude solitons $(E_m 
\ll 1)$ the soliton represents a soliton solution of the cubic 
nonlinear Schr\"{o}dinger equation and can be written as:
$$E = E_m sech \left [ \left({3\over 2}\right)^{1/2} E_m z \right] 
\eqno(25) $$
If $E_m \rightarrow 1$ $(\lambda^2 \rightarrow 1.2)$ the top 
part of the soliton is well described by a $cosine$ function and can be 
approximated as 
$$E = E_m cos[(2-\lambda^2)^{1/2}z]  \eqno (26)$$
The general shape of the soliton is displayed in Fig.1 where 
$E_m=0.99$ $(\lambda^2=1.19)$. Dashed line corresponds to analytical 
approximation given by Eq.(26). 

Using the well-known stability criterion 
of Vakhitov and Kolokolov [9], it can be shown that the above described 
soliton-type solution is stable against small perturbations.

In conclusion, we have considered the possibility of high-frequency EM wave 
localization in hot unmagnetized electron-positron plasmas. In our 
analysis, we included not only the relativistic effects in the 
hydrodynamic motion of the plasma, but also the effects which result from 
the relativistic electron velocity distribution. We have shown that in 
such plasmas it is possible to have localized stable soliton-like 
structures. It is also shown that cavitation of plasma can occur both in
the nonrelativistic and ultrarelativistic cases.
The present result should be useful for the understanding of the 
nonlinear photon motion in cosmical plasmas such as those found in the 
early universe and AGN.
\vskip 1.0truecm
This work was supported, in part, by ISF grant No. KZ3200. 
The work of S. K. was supported by the Bogazici University 
Center for Turkish-Balkan Physics Research and Applications (ICBP)
and the Turkish Scientific and Technological Research Council  
(TUBITAK).
\newpage
\centerline{\bf References}
\begin{description}

\item{[1]}
C.F. Kennel and R. Pellat, J. Plasma Phys. {\bf 15}, 335 (1976);\\
J.N. Leboeuf, M. Ashour-Abdalla, T. Tajima, C.F. Kennel, F. Coroniti,
and J.M. Dawson, Phys. Rev. A{\bf 25}, 1023 (1982);
M.E. Gedalin, J.G. Lominadze, L. Stenflo, and V.N. Tsitovich,
Astrophys. Space Sci. {\bf 108}, 393 (1985);
P.K. Shukla, N.N. Rao, M.Y. Yu, and N.L. Tsintsadze, Phys. Rep. {\bf 135},
1 (1986);
V.I. Berezhiani, V. Skarka, and S.M. Mahajan, Phys. Rev. E{\bf 48}, R3252 (1993).
\item{[2]}
F.C. Michel, Rev. Mod. Phys. {\bf 54}, 1 (1982);
M.C. Begelman, R.D. Blandford, and M.D. Rees, Rev. Mod. Phys. {\bf 56},
255 (1984).
\item{[3]}
K. Holcomb and T. Tajima, Phys. Rev. D{\bf 40}, 3909 (1989);
T. Tajima and T.Taniuti, Phys. Rev. A{\bf 42}, 3587 (1990);
P.K. Shukla, N.L. Tsintsadze, and L.N. Tsintsadze, Phys. Fluids B{\bf 5},
233 (1993).
\item{[4]}
E. Witten, Nucl. Phys. B{\bf 249}, 557 (1985);
J.P. Ostriker, C. Thompson, and E. Witten, Phys. Lett. B{\bf 180}, 231 (1986).
\item{[5]}
A.C.L. Chian and C.F. Kennel, Astrophys. Space Sci. {\bf 97}, 3 (1983);
A.B. Mikhailovskii, O.G. Onishchenko, and E.G. Tatarinov, Plasma Phys. Contr.
Fusion, {\bf 27}, 539 (1986);
R.T. Gangadhara, V. Krishan, and P.K. Shukla, Mon. Not. R. astr. Soc. {\bf 258},
616 (1992).
\item{[6]}
V.I. Berezhiani and S.M. Mahajan, Phys. Rev. Lett. {\bf 73}, 1110 (1994);
V.I. Berezhiani and S.M. Mahajan, Phys. Rev E{\bf 52}, 1968 (1995).
\item{[7]}
D.I. Dzhavakhishvili and N.L. Tsintsadze, Sov. Phys. JETP {\bf 37},
666 (1973);
S.V. Kuznetsov, Sov. J. Plasma Phys. {\bf 8}, 199 (1982).
\item{[8]}
L.N. Tsintsadze, Phys. Scr. {\bf 50}, 413 (1994).
\item{[9]}
J. Juul Rasmussen and K. Rypdal, Phys. Scr. {\bf 33}, 481 (1986).

\end{description}
\newpage
\centerline{\bf Figure Captions}
\vskip 1.5truecm
Fig.1  Solution $E$ as a function of the space coordinate $z$, 
$(E_m=0.99)$. The dashed line corresponds to analytical approximation 
given by Eq.(26).

\end{document}